\documentclass[showpacs,amsmath,amssymb,pre,preprint]{revtex4-1}
\usepackage{graphicx}
\usepackage{bm}

\begin{document}

\title{Harmonic analysis of irradiation asymmetry for cylindrical implosions driven by
high-frequency rotating ion beams}

\author{A. Bret}\author{A.R. Piriz}
\affiliation{ETSI Industriales, Universidad de Castilla-La Mancha, 13071 Ciudad Real, Spain}
 \affiliation{Instituto de Investigaciones Energ\'{e}ticas y Aplicaciones Industriales,
 Campus Universitario de Ciudad Real,
 13071 Ciudad Real, Spain.}

 \author{N. Tahir}
\affiliation{GSI Darmstadt, Plankstrasse 1, 64291 Darmstadt, Germany}

\begin{abstract}
Cylindrical implosions driven by intense heavy ions beams should be instrumental in a near future to study High Energy Density Matter. By rotating the beam by means of a high frequency wobbler, it should be possible to deposit energy in the outer layers of a cylinder, compressing the material deposited in its core. The beam temporal profile should however generate an inevitable irradiation asymmetry likely to feed the Rayleigh-Taylor instability (RTI) during the implosion phase. In this paper, we compute the Fourier components of the target irradiation in order to make the junction with previous works on RTI performed in this setting. Implementing a 1D and 2D beam models, we find these components can be expressed exactly in terms of the Fourier transform of the temporal beam profile. If $T$ is the beam duration and $\Omega$ its rotation frequency, ``magic products'' $\Omega T$ can be identified which cancel the first harmonic of the deposited density, resulting in an improved irradiation symmetry.
\end{abstract}


\maketitle

\section{Introduction}
The study of matter under extreme conditions of pressure and density is of great interest
for astrophysics, planetary sciences and inertial fusion \cite{Guillot1999,NettelmannApJ,drake,tahirNJP}. Particularly appealing is the perspective to produce cylindrical implosions with a high degree of symmetry. Such experiments would allow to generate large volumes of high-energy density matter (HEDM), including strongly coupled plasmas. Among others, the experimental realization of the once predicted  metallic state of Hydrogen \cite{wigner} is envisioned at the Gesellschaft f\"{u}r Schwerionenforschung (GSI) near Darmstadt, Germany. Within the framework of the Facility for Antiproton and
Ion Research (FAIR) currently under construction at GSI \cite{Henning}, the so-called LAPLAS experiment (LAboratory of PLAnetary Sciences) aims at implementing such cylindrical scheme to study equation of state and transport properties of HEDM \cite{Tahir2000a,Tahir2000b,Tahir2001}.

Figure 1 sketches the typical experimental scheme implemented for cylindrical implosions. A hollow ion beam hits the absorber part  of the cylinder. By tailoring its energy so that the Bragg peak of the ions falls well outside the cylinder, it is possible to obtain a very homogenous energy deposition inside the cylinder. The following compression has been the object of various numerical and analytic studies in recent years, to assess, among other, the impact of the Rayleigh-Taylor instability (RTI) on the medium interfaces during the compression \cite{PirizPRE2002,PirizPPCF,PirizPRE2003,Basko2004,PirizPRE2009,tahirNJP}.

The hollow beam generation should be achieved by a high frequency wobbler rotating the ion beam \cite{Sharkov2001}. The rotation frequency necessary to achieve an acceptable degree of irradiation symmetry has been calculated by Piriz \emph{et al.} \cite{PirizPRE2003} and found in the GHz range. For such a fast deposition, the target motion during the deposition can be neglected. But even so, an inevitable source of asymmetry comes from the beam temporal profile itself. Because the beam duration is finite, the number of ions deposited along the absorber will  be inhomogeneous.  The overall effect of such asymmetry on the pressure driving the compression has been investigated in Refs. \cite{PirizPPCF} and \cite{Basko2004}, where the amplitude of the asymmetry has been evaluated. But previous works on the RTI for this setting have showed that stability is a matter of both the excited wavelength, and the excitation amplitude. It is thus necessary to know the Fourier components of the ion deposition in the pusher, precisely because RTI analysis is performed in Fourier space. This procedure is reminiscent of studies conducted for Inertial Confinement Fusion, where the driver energy deposition on the spherical pellet has to be Fourier analyzed to assess its RT stability \cite{Lindl1995,Logan2009}.

The goal of this paper is to compute the Fourier components of the ion deposition in the absorber. We start implementing a 1D beam model and express the harmonics amplitude of the ion density deposition, in terms of the Fourier transform of the temporal beam profile. We then consider a beam spatially extended in the transverse direction, and generalize the 1D results. We then turn to the stability analysis, making the junction with previous works. Finally, we address the possibility of canceling the first harmonic. In the case of a parabolic temporal profile, an infinite series of product $\Omega T$ are found for which the first harmonic vanishes exactly.

\begin{figure}[t]
\includegraphics[width=0.5\textwidth]{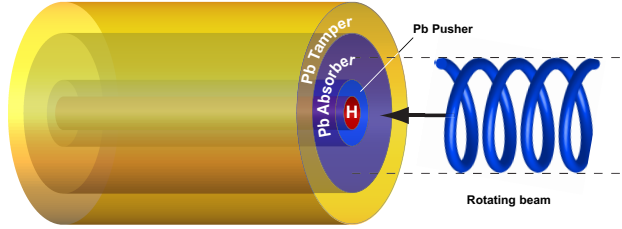}
\caption{(Color online) Typical cylindrical implosion scheme. Not to scale.} \label{fig1}
\end{figure}

\section{1D beam model}
\subsection{Irradiation spectrum calculation}
\begin{figure}[t]
\includegraphics[width=0.9\textwidth]{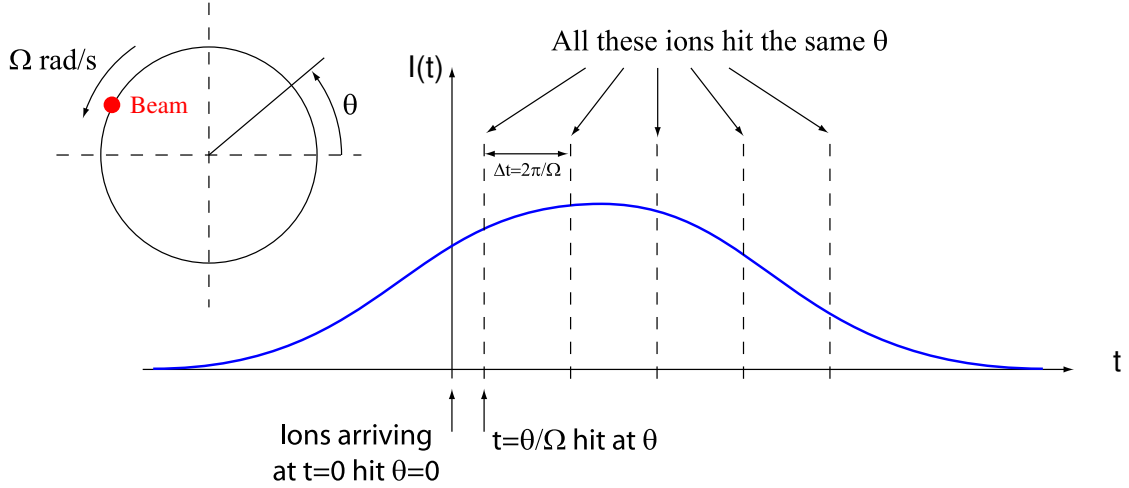}
\caption{(Color online) The beam intensity profile is $I(t)$. It rotates at
velocity $\Omega$ rad/s along the circle. Ions arriving at
$t=\theta/\Omega + 2k\pi/\Omega$ all hit the same point of the
circle.} \label{fig2}
\end{figure}

Consider a 1D ion beam with temporal profile $I(t)$, normalized to unity so that $\int I(t)dt=1$. The beam is rotated by a wobbler over the absorber region at velocity $\Omega$ rad/s (see Figure \ref{fig2}). The number of ions brought by the beam between $t$ and $t+dt$ is $I(t)dt$. Suppose ions arriving at $t=0$ hit the target at $\theta=0$. Then those arriving at
$t=\theta/\Omega$ hit the target at angle $\theta$, as explained on the Figure. The number of ions deposited between $\theta$ and $\theta+d\theta$, from time $t$ to time $t+dt$ is therefore,
\begin{equation}\label{eq:1}
    dn(\theta)=\lim_{d\theta\rightarrow 0}\int_{\frac{\theta}{\Omega}}^{\frac{\theta+d\theta}{\Omega}}I(u)du=I\left(\frac{\theta}{\Omega}\right)\frac{d\theta}{\Omega}
\end{equation}
Then, ions arriving $2l\pi/\Omega$ seconds
later, or before, hit the very  same point, for any $l$ integer, positive or
negative. The total number of ions eventually deposited between $\theta$ and $\theta+d\theta$
is thus
\begin{eqnarray}\label{eq:N}
   dN(\theta)&=&\sum_{l=-\infty}^\infty I\left(\frac{\theta}{\Omega}+\frac{2l\pi}{\Omega}\right)\frac{d\theta}{\Omega} \nonumber \\
   \Rightarrow
    \frac{dN(\theta)}{d\theta} &=&\frac{1}{\Omega} \sum_{l=-\infty}^\infty  I\left(\frac{\theta+2l\pi}{\Omega}\right)\equiv\rho(\theta).
\end{eqnarray}
As expected, the density deposition $dN(\theta)/d\theta\equiv\rho(\theta)$ is periodic, of period $2\pi$.

We now turn to the Fourier transform of the density deposition $\rho(\theta)$ in the absorber ring,
\begin{eqnarray}\label{eq:Fourier}
   \widehat{\rho}(s)&=&\int_{-\infty}^\infty\rho(\theta)e^{is\theta}d \theta
          =\frac{1}{\Omega}\int_{-\infty}^\infty\sum_{l=-\infty}^\infty I\left(\frac{\theta+2l\pi}{\Omega}\right)e^{is\theta}d\theta\nonumber\\
    &=&\frac{1}{\Omega}\sum_{l=-\infty}^\infty\int_{-\infty}^\infty I\left(\frac{\theta+2l\pi}{\Omega}\right)e^{is\theta}d\theta.
\end{eqnarray}

Having a periodic function, it could be possible to Fourier analyze it integrating from 0 to $2\pi$. However, integrating from $-\infty$ to $\infty$ is more general and allows to deal with the Fourier Transform of the beam profile as well. We may now permute the infinite sum with the integral because we are dealing with ``real world'' regular functions for which such operation can be done. Making the substitution $u =
\theta/\Omega + 2l\pi/\Omega$, we find
\begin{eqnarray}\label{eq:Fourier2}
   \widehat{\rho}(s)
    &=&\sum_{l=-\infty}^\infty\int_{-\infty}^\infty
    I(u)e^{isu\Omega-2il\pi s} du=\sum_{l=-\infty}^\infty\int_{-\infty}^\infty
    I(u)e^{is\Omega u}e^{-2il\pi s} du \nonumber\\
    &=&\sum_{l=-\infty}^\infty\left(\int_{-\infty}^\infty
    I(u)e^{is\Omega u} du\right)e^{-2il\pi s}.
\end{eqnarray}
The term between parenthesis is just the Fourier transform of the
beam temporal profile for the frequency value ``$s\Omega$''. We thus
have
\begin{equation}\label{eq:Fourier3}
   \widehat{\rho}(s)
    =\sum_{l=-\infty}^\infty \widehat{I}(s\Omega) e^{-2il\pi
    s}=\widehat{I}(s\Omega)\sum_{l=-\infty}^\infty  e^{-2il\pi
    s}.
\end{equation}
According to some properties of the ``Dirac's comb'' function $\sum  \delta(s-l)$ \cite{bracewell}, we have
\begin{equation}\label{eq:peigne}
 \sum_{l=-\infty}^\infty  e^{-2il\pi s} =  \sum_{l=-\infty}^\infty  \delta(s-l),
\end{equation}
and we finally come to,
\begin{equation}\label{eq:Fourier4}
   \widehat{\rho}(s)
    =\sum_{l=-\infty}^\infty \widehat{I}(s\Omega) e^{-2il\pi
    s}=\widehat{I}(s\Omega) \sum_{l=-\infty}^\infty
 \delta(s-l).
\end{equation}
The infinite sum $\sum\delta(s-l)$ functions at the right hand
side implies the spectrum is discrete. This comes
from the fact that we consider the Fourier transform of the  periodic
function $\rho(\theta)$. Furthermore, the sum of $\delta$'s implies that the
discrete values of the spectrum are the integers
$l=-\infty\ldots\infty$. The amplitudes of the harmonics are therefore,
\begin{eqnarray}\label{eq:Resultat}
  \widehat{\rho}(0) &=& \widehat{I}(0), \nonumber\\
  \widehat{\rho}(1) &=& \widehat{I}(\Omega) + \widehat{I}(-\Omega), \nonumber\\
  \widehat{\rho}(2) &=& \widehat{I}(2\Omega)+ \widehat{I}(-2\Omega), \nonumber\\
  &\cdots& \nonumber\\
  \widehat{\rho}(l) &=& \widehat{I}(l\Omega)+ \widehat{I}(-l\Omega), ~~\forall l\in \mathbb{N}.
\end{eqnarray}
From the harmonics above, the density deposition $\rho(\theta)$ is expressed as
\begin{equation}\label{eq:FourierInv}
    \rho(\theta)=\frac{1}{2\pi}\sum_{l=-\infty}^\infty\widehat{\rho}(l)e^{-il\theta}
    =\frac{1}{2\pi}\sum_{l=-\infty}^\infty\widehat{I}(l\Omega)e^{-il\theta},
\end{equation}
with
\begin{equation}\label{eq:FourierI(t)}
    \widehat{I}(\omega)=\int_{-\infty}^\infty I(t)e^{i\omega t} dt.
\end{equation}

Such is the final result, where the amplitude of the harmonics of the density deposition is expressed in terms of the beam temporal profile. Let us now illustrate the result considering the example of a parabolic temporal profile.

\subsection{Parabolic beam temporal profile}
\begin{figure}[t]
\includegraphics[width=0.5\textwidth]{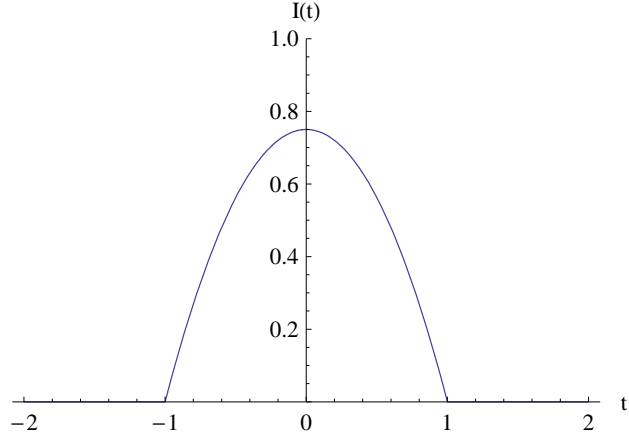}
\caption{Temporal profile of the beam from Eq. (\ref{eq:para}) with $T=2$ s. The number of ions hitting the target from $t$ to $t+dt$ is $I(t)dt$. $\int I(t)dt=1$.} \label{fig:I(t)}
\end{figure}

\begin{figure}[t]
\includegraphics[width=0.5\textwidth]{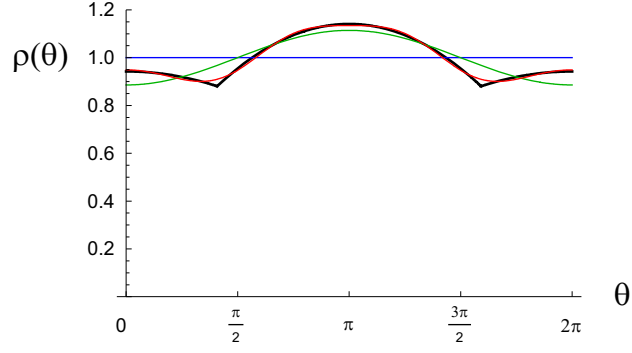}
\caption{(Color online) Black: Density profile calculated from Eq. (\ref{eq:1}). Blue: Term $l=0$ in Eq. (\ref{eq:FourierInv}). Green: Eq. (\ref{eq:FourierInv}) up to $l=\pm 1$. Red: Eq. (\ref{eq:FourierInv}) up to $l=\pm 4$. Parameters are $T=2$ s and $\Omega$ = 5 rad/s.} \label{fig:fourier}
\end{figure}

For a  parabolic temporal beam profile normalized to 1 (see Fig. \ref{fig:I(t)}),
\begin{eqnarray}\label{eq:para}
    I(t) &=&
    \frac{3}{2T}\left[1-\left(\frac{t}{T/2}\right)^2\right],~\mathrm{for} -T/2<t<T/2
    \nonumber\\
    &=&0 ~~\mathrm{otherwise},
\end{eqnarray}
we find for the $n^{th}$ harmonic of the density deposition $\rho(\theta)$,
\begin{equation}\label{eq:harmopara}
     \widehat{\rho}(n)=\widehat{I}(n\Omega)+  \widehat{I}(-n\Omega)= \frac{-12\cos (nT\Omega/2)}{(nT\Omega)^2} +
  \frac{24\sin(nT\Omega/2)}{(nT\Omega)^3}.
\end{equation}
Figure \ref{fig:fourier} shows the density $\rho(\theta)$ profile calculated from Eq. (\ref{eq:N}) and the terms $l=0$ of Eq. (\ref{eq:FourierInv}) together with  sum (\ref{eq:FourierInv}) up to $l=\pm 1$ and $\pm 4$. The term $l=0$ is just the mean value of the density deposition. One can notice how the sum up to $l=\pm 4$ is already very close to the exact density.

Let $R$ be the radius of the circle where the ion beam are
deposited, the $n^{\mathrm{th}}$ harmonic pertains to the wavelength
$\lambda = 2\pi R/n$ and the wave vector
\begin{equation}\label{eq:k}
k=\frac{2\pi}{\lambda}=\frac{n}{R}.
\end{equation}
Let us assume the wobbler rotates $N$ times in $T$ seconds, that is
$T\Omega = 2\pi N$, the relative amplitude of the first harmonic
is
\begin{equation}\label{eq:amplirelat}
    \frac{\widehat{\rho}(1)}{\widehat{\rho}(0)} = \frac{3}{\pi^2 N^2}=0.3\%~\mathrm{for} ~N=10.
\end{equation}

\subsection{General analysis for finite duration beams}
We now derive some general result regarding functions $I(t)$ which vanish out of a finite time range $[0,T]$.
According to Eq. (\ref{eq:Resultat}), the $n^{\mathrm{th}}$ harmonic amplitude of the
ion density deposition is given
\begin{equation}\label{eq:fini1}
    \widehat{\rho}(n) =  \widehat{I}(n\Omega)+\widehat{I}(-n\Omega)=2\int_{-\infty}^\infty I(t)\cos(n\Omega t)dt=2\int_0^T I(t)\cos(n\Omega t)dt.
\end{equation}
Integrating by part, we find
\begin{equation}\label{eq:fini2}
    \widehat{\rho}(n) = \frac{2}{n \Omega}\left[I(t)\sin(n\Omega t)\right]_0^T - \frac{2}{n \Omega}\int_0^T I'(t)\sin(n\Omega t)dt,
\end{equation}
where $I'(t)=dI/dt$. Repeating again the process gives,
\begin{equation}\label{eq:fini3}
    \widehat{\rho}(n) = \frac{2}{n \Omega}\left[I(t)\sin(n\Omega t)\right]_0^T +\frac{2}{n^2\Omega^2 }\left[I'(t)\cos(n\Omega t)\right]_0^T - \frac{1}{n^2\Omega^2 }\int_0^T I''(t)\cos(n\Omega t)dt.
\end{equation}
It is thus straightforward that if $I(0)=I(T)=0$, while the first
derivative does not vanish in $t=0$ or $T$, $\widehat{\rho}(n)$ behaves like
$1/n^2$. More generally, $\widehat{\rho}(n)$ behaves like $1/n^{k+1}$ if all
the derivative up to the $k^{th}$ vanish in $t=0$ and $T$. We thus
come to the same conclusion than Ref. \cite{PirizPPCF}. Additionally, we find that harmonic amplitudes can be tuned by choosing $\Omega T$ appropriately. This important point is detailed in Section \ref{sec:cancel}.

\section{2D beam model}
\subsection{Irradiation spectrum calculation}
We have so far considered a point-like beam with a temporal
profile. We thus need to account for a beam with spatial
extension. The beam transverse density is defined in its own rest frame by
\begin{equation}\label{eq:real1}
    I(t,\mathbf{u})=\xi(t)\sigma(\mathbf{u})=\xi(t)\sigma(u),
\end{equation}
where $u$ is the distance from the beam center to a given beam point. We thus consider beams with cylindrical symmetry.

\begin{figure}[t]
\includegraphics[width=0.6\textwidth]{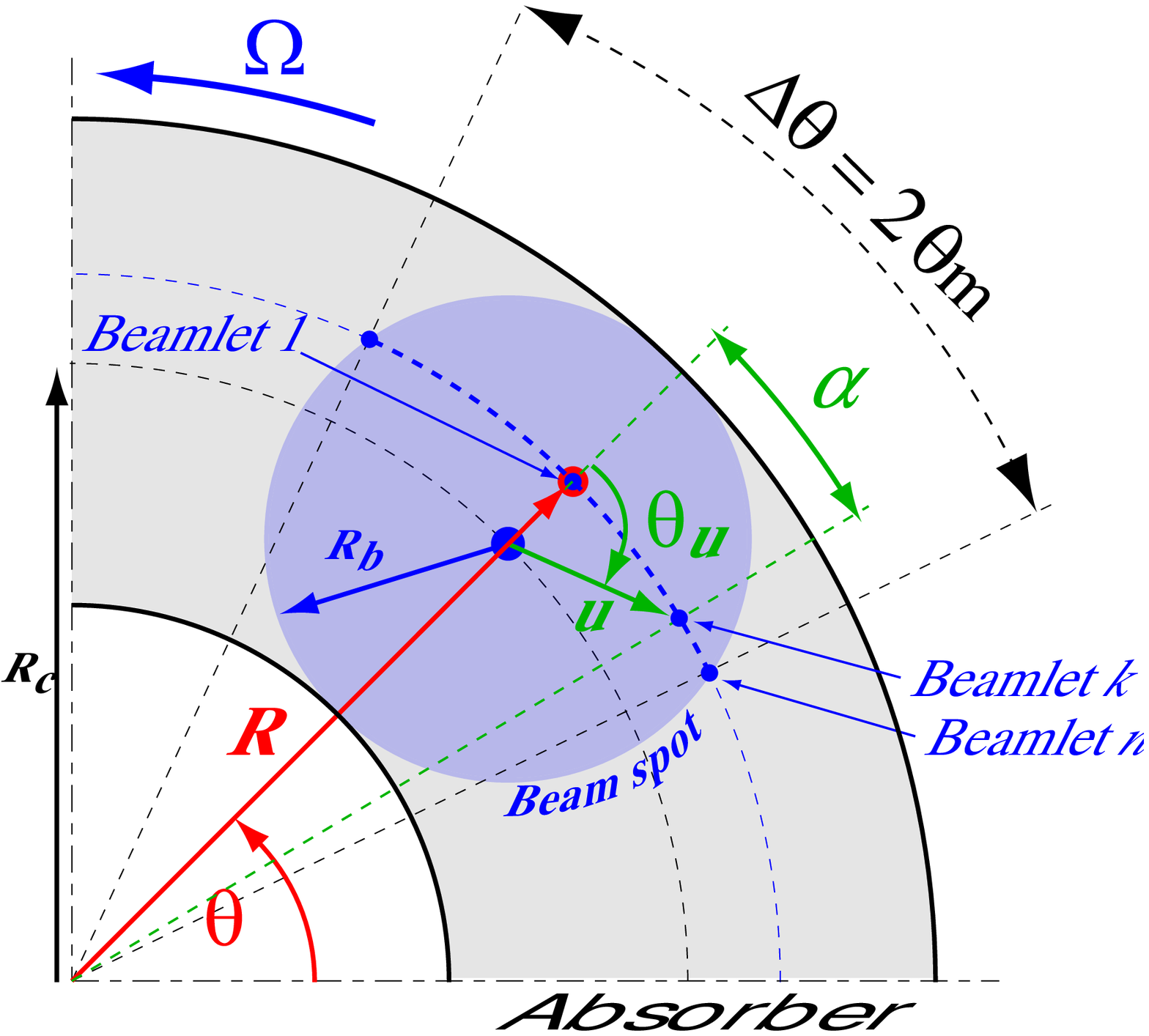}
\caption{(Color online) Calculation for a beam with temporal and transverse spatial profile. Not to scale.} \label{fig6}
\end{figure}

Let us calculate the total number of ions arriving at the target point localized
by $(R,\theta)$ (the red point on Fig. \ref{fig6}). We can
consider here that the target is hit by various beamlets. The target
point $(R,\theta)$ is hit by $n$ ``beamlets'' represented by the
blue circles. Of course, $n$ is eventually infinite. We thus start the analysis with $n$ finite, before we make it tend to infinity.

Reasoning first in the target polar coordinates (the green elements on Fig. \ref{fig6}), the number of ions brought by the first beamlet is,
\begin{equation}
dN=\sigma(u_1) u_1d\theta_u ~ du
\end{equation}
By analogy with Eq. (\ref{eq:N}) from the 1D calculation, we deduce the ions density deposited by this beamlet,
\begin{equation}\label{eq:real2}
    d\rho_1(R,\theta)=\frac{1}{\Omega}\sum_{l=-\infty}^\infty
    \xi\left(\frac{\theta+2l\pi}{\Omega}\right)\times \sigma(u_1) u_1d\theta_u du.
\end{equation}
All the beamlets pictured on Fig. \ref{fig6} up to ``beamlet $n$'' will deposit ions in $(R,\theta)$. But it is clear that beamlets with $n>1$ deposit their ions at target point $(R,\theta)$ some time later, depending on how large is the angle $\alpha$ the beam needs to rotate for those beamlets to cover the target point. Therefore, the beamlet number $k$, deposits
\begin{equation}\label{eq:real3}
    d\rho_k(R,\theta)=\frac{1}{\Omega}\sum_{l=-\infty}^\infty
    \xi\left(\frac{\theta+\alpha(u,\theta_{u})+2l\pi}{\Omega}\right)\sigma(u) u d\theta_u du.
\end{equation}
Summing the contributions from all beamlets, we integrate over the arc of radius $R$ intersecting the beam spot. It is of course much more convenient to work in polar coordinates with respect to the target center instead of the beam spot center. We thus integrate the expression above from $\alpha=\theta_m$ to $\theta_m$. When switching from $(u,\theta_u)$ to $(R,\alpha)$ coordinates, the differential element $d\theta_u du$ changes to $J(R,\theta)drd\theta$, where $J$ is the Jacobian of the transformation. We thus find

\begin{equation}\label{eq:real4}
\rho(R,\theta)=\frac{1}{\Omega}\int_{\alpha=-\theta_m}^{\theta_m}\sum_{l=-\infty}^\infty\xi\left(\frac{\theta+\alpha+2l\pi}{\Omega}\right)\sigma\left[u(R,\alpha)\right]u(R,\alpha) J(R,\alpha)d\alpha~dr.
\end{equation}
This result simplifies substantially due to the identity proven in Appendix \ref{sec:ApA}
\begin{equation}\label{eq:real5}
    u(R,\alpha)J(R,\alpha)=R,
\end{equation}
so that
\begin{equation}\label{eq:real6}
\rho(R,\theta)=
\frac{R dr}{\Omega}\int_{\alpha=-\theta_m}^{\theta_m}\sum_{l=-\infty}^\infty\xi\left(\frac{\theta+\alpha+2l\pi}{\Omega}\right)
\sigma\left[u(R,\alpha)\right]  d\alpha.
\end{equation}
Note that $\rho(R,\theta)/Rdr$ can be interpreted as the ion density deposited in the absorber along the circle on radius $R$. Also, integration from $\alpha=-\theta_m$ to $\theta_m$ can be replaced by an integration from $-\infty$ to $\infty$ since the integrand vanishes out of $[-\theta_m,\theta_m]$. Our interest now lies in the Fourier transform with respect to $\theta$ of the function above. Noteworthily, Eq. (\ref{eq:real6}) can be written as,
\begin{equation}\label{eq:conv0}
    (f*g)(\theta)=\int_{-\infty}^\infty f(\theta+\alpha)g(\alpha)d\alpha,
\end{equation}
with
\begin{eqnarray}
  f(\tau) &=& \frac{R dr}{\Omega}\sum_{l=-\infty}^\infty \xi\left(\frac{\tau+2l\pi}{\Omega}\right), \nonumber\\
  g(\tau) &=& \sigma\left[u(R,\tau)\right].
\end{eqnarray}
Changing $\alpha \rightarrow -\alpha$, and noting that $g$ is an even function of $\alpha$, we rewrite  Eq. (\ref{eq:conv0}),
\begin{equation}\label{eq:conv1}
    (f*g)(\theta)=\int_{-\infty}^\infty f(\theta-\alpha)g(\alpha)d\alpha,
\end{equation}
which appears to be the convolution product of the two functions.
The first of these function is very reminiscent of Eq. (\ref{eq:N}) from the 1D case. It Fourier transform with respect to $\theta$ can be calculated following the very same line and is found adapting Eq. (\ref{eq:Fourier4}) as
\begin{equation}\label{eq:Fourier4_2D}
R dr\widehat{\xi}(s\Omega) \sum_{l=-\infty}^\infty \delta(s-l).
\end{equation}
Because the Fourier transform of a convolution product is the product of the Fourier transforms, we can now write directly
\begin{equation}\label{eq:Fourier4_OK}
\frac{\widehat{\rho}(R,s)}{R dr}\equiv \widehat{\rho}_{2D}(R,s) = \left[\widehat{\xi}(s\Omega) \sum_{l=-\infty}^\infty \delta(s-l)\right]\widehat{\sigma}\left[u(R,s)\right].
\end{equation}

The interpretation of the equation above is simple: the $n^{th}$
harmonic with respect to $\theta$ is now the $n^{th}$ harmonic of the
point-like beam, times the $n^{th}$ harmonic of a beam form factor. For a point-like beam,
this factor is a $\delta$ function which Fourier transform
is 1. This equation also shows that one should expect greater
homogeneity with a spread out beam since both harmonics (temporal
and spatial) are multiplied to yield the deposition harmonics.

\subsection{Parabolical beam in time and space}
\begin{figure}[t]
\includegraphics[width=0.6\textwidth]{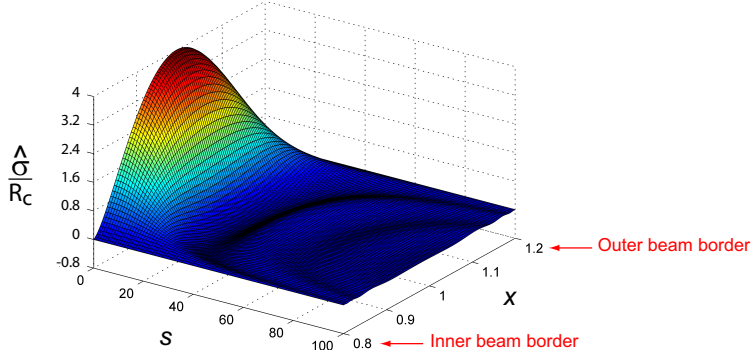}
\caption{(Color online) Plot of Equation (\ref{eq:fourier_2D1}) for $\epsilon=1/5$. The inner border of the beam spot is defined by $x=0.8=1-\epsilon$, and the outer border by $x=1.2=1+\epsilon$.} \label{fig:7}
\end{figure}

We now apply the formalism developed previously to a
parabolical beam both in time and space. The temporal profile
$\xi(t)$ is thus identical to the function defined by Eq.
(\ref{eq:para}),
\begin{eqnarray}\label{eq:parareal1}
    \xi(t) &=&
    \frac{2}{3T}\left[1-\left(\frac{t}{T/2}\right)^2\right],~\mathrm{for} -T/2<t<T/2
    \nonumber\\
    &=&0 ~~\mathrm{otherwise},
\end{eqnarray}
and its harmonics are
\begin{equation}\label{eq:parareal2}
   \widehat{ \xi}(s) = 12\frac{   2\sin(sT/2)-sT\cos(sT/2)  }{s^3T^3}.
\end{equation}
The spatial profile $\sigma(u)$ is
\begin{eqnarray}
    \sigma(u) &=&
    \frac{2}{\pi R_b^2}\left[1-\left(\frac{u}{R_b}\right)^2\right],~\mathrm{for} -R_b<u<R_b
    \nonumber\\
    &=&0 ~~\mathrm{otherwise}.
\end{eqnarray}
Choosing here $\theta=(\widehat{\mathbf{R}_c,\mathbf{R}})$, the vectorial relation $\mathbf{R}_c+\mathbf{u}=\mathbf{R}$ allows to derive,
\begin{equation}\label{eq:u}
u(R,\theta)=\sqrt{R^2-R_c^2-2R_cu\cos\theta},
\end{equation}
and,
\begin{equation}\label{eq:sigma}
    \sigma[u(R,\theta)]=2\frac{R_b^2 - R^2 - R_c^2 + 2r R_c\cos\theta }{\pi R_b^4},
\end{equation}
with $r\in [R_c-R_b,R_c+R_b]$ and $\theta\in [-\theta_m,\theta_m]$ with
\begin{equation}\label{eq:sigma2}
\cos\theta_m=\frac{R^2+R_c^2-R_b^2}{2RR_c}.
\end{equation}
The Fourier transform of $\sigma[u(R,\theta)]$ is
\begin{equation}\label{eq:fourier_2D}
    \widehat{\sigma}[u(R,s)] = \int_{-\infty}^\infty\sigma[u(R,\theta)]e^{is\theta}d\theta
    =\int_{-\theta_m}^{\theta_m}\sigma[u(R,\theta)]e^{is\theta}d\theta .
\end{equation}
The result reads,
\begin{equation}\label{eq:fourier_2D1}
    R_c^2~\widehat{\sigma}[u(x,s)]=
    -4\frac{ s x \sqrt{\frac{2 x^2 \left(1+\epsilon ^2\right)-x^4-(\epsilon ^2-1)^2}{x^2}} \cos(s \Theta)+\left(\epsilon ^2-1-x^2\right) \sin(s\Theta)}{\pi R_c^2 s (s^2-1) \epsilon ^3},
\end{equation}
where,
\begin{equation}
    x=\frac{R}{R_c}\in [1-\epsilon,1+\epsilon],~~\epsilon=\frac{R_b}{R_c},~~\Theta=\arccos\left[\frac{1+x^2-\epsilon ^2}{2 x}\right].
\end{equation}
Each circle of radius $r$ thus yields a different Fourier transform. Equation (\ref{eq:fourier_2D1}) is plotted on Figure \ref{fig:7} for $\epsilon=1/5$. The spectrum is continuous in $s$ and of infinite extension because the source function is finite only over a finite $\theta$ interval. Because ion deposition tends to zero near the borders of the beam, the harmonics amplitude vanish for $x=1\pm\epsilon$.

\subsection{1D approximation}
Upon which condition shall the Fourier transform in the 2D model be comparable to the 1D one, for any radius? If the form factor $\widehat{\sigma}[u(R,s)]$ remains constant over the first harmonics of the temporal profile. Indeed, these first harmonics are the most important, because their amplitude is larger. By property of the Fourier transform, the function  $\widehat{\sigma}[u(R,s)]$ has an extension $\Delta s$ such that $\Delta s\Delta \theta\sim 1$, where $\Delta\theta$ is the extension of $\sigma[u(R,\theta)]$. It is clear that maximal extension is achieved for $R\sim R_c$, with $\Delta\theta\sim R_b/R_c$. Hence, the Fourier transform $\widehat{\sigma}[u(r,s)]$ has a minimal extension $\sim R_c/R_b$. Confirmation if found on Fig. \ref{fig:7}, where $\widehat{\sigma}[u(x,s)]$ is all the more peaked than $x$ is close to 1, i.e. $R=R_c$.

Therefore, if $\widehat{\sigma}[u(R,s)]$ can be considered constant for $s\lesssim R_c/R_b$ while the 1D model has its harmonic for $s=1,2,\ldots$, then the beam can be considered point-like for harmonics $s=1\ldots R_c/R_b$, implying only $R_c>R_b$. With $R_c=5R_b$ for example, harmonics up to 5 should be close to the 1D model even at $R=R_c$.

Turning now to the stability analysis, we use the 1D model, precisely because the most relevant harmonics are the first ones. Indeed, we will check that the analysis can be conducted in terms of the $s=1$ harmonic only. For the 1D model to be valid, we thus simply need $R_b<R_c$.

\section{Stability analysis}
The RTI at the absorber-pusher interface has been studied in Ref. \cite{PirizPRE2009}. Due to the expected experimental conditions, both sides of the interface are occupied by solids where elastic-plastic effects have to be accounted for. In this respect, it has been found that the RTI for a perturbation of wavelength $\lambda$ and initial amplitude $\xi_0$ depends on the two dimensionless parameters,
\begin{eqnarray}\label{eq:dimless}
  \widehat{\lambda} &=& \frac{\rho g \lambda}{4\pi G}, \nonumber\\
  \widehat{\xi} &=& \frac{\rho g \xi_0}{\sqrt{3}Y},
\end{eqnarray}
where $g$ is the acceleration of the interface, $\rho$, $G$ and $Y$ the density, the shear modulus and the yield strength of the pusher respectively.

\begin{figure}[t]
\includegraphics[width=0.8\textwidth]{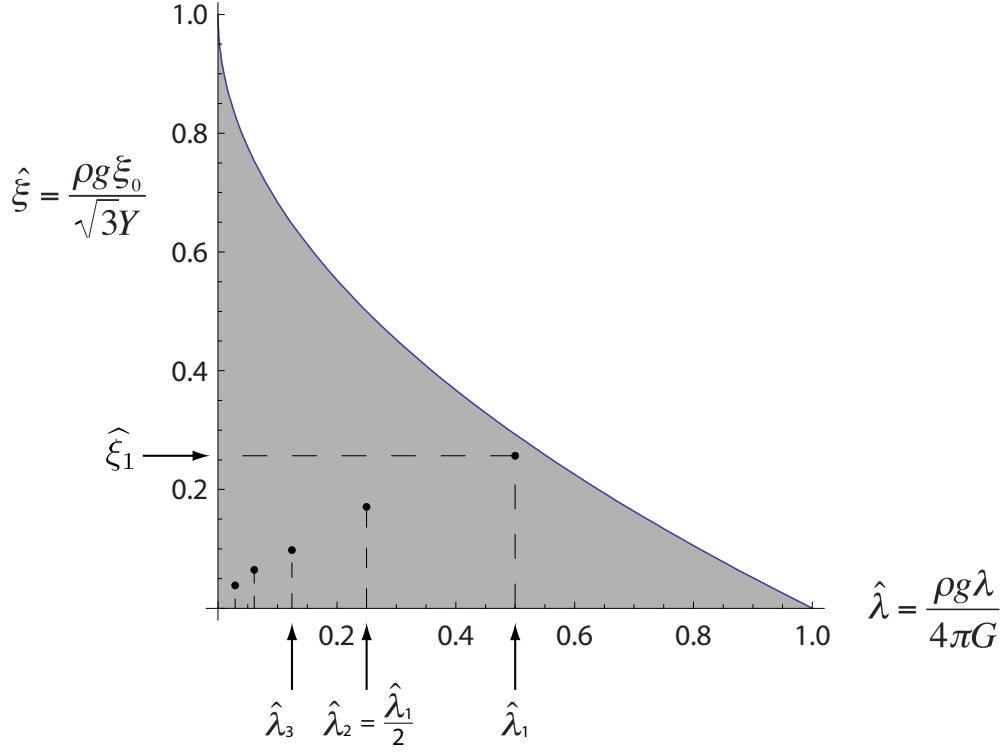}
\caption{Stability domain - shaded area -  in terms of the dimensionless parameters $\widehat{\lambda}$ and $\widehat{\xi}$. If $\lambda_1$ is stable, then $\lambda_n$ is stable $\forall n$. From Ref. \cite{PirizPRE2009}.} \label{fig:stab}
\end{figure}

The stability diagram  appears on Fig. \ref{fig:stab}. In the present problem, the excited wave-lengths are $\lambda_n=2\pi R/n$, $n\in \mathbb{N}$. The $\lambda$'s are thus decreasing with $n$, the largest one being $\lambda_1=2\pi R$. For this mode to be stable, i.e. to lie in the shaded area on Fig. \ref{fig:stab}, it is \emph{necessary} (but not \emph{sufficient}) that
\begin{equation}\label{eq:nece}
    \widehat{\lambda}_1=\frac{\rho g \lambda_1}{4 \pi G}<1\Rightarrow  \frac{\rho g R}{2 G}<1.
\end{equation}

In the stability diagram, the excited modes yield points $(\widehat{\lambda}_n,\widehat{\xi}_n)$. On the one hand, it is clear that $\widehat{\lambda}_n<\widehat{\lambda}_1$ for $n\neq 1$. On the other hand, harmonics amplitude are usually decreasing functions of $n$. This can be checked on Eq. (\ref{eq:harmopara}) for the parabolic time profile beam, or more generally on Eq. (\ref{eq:fini3}). To be more accurate, we can say that the harmonics as a function of $n$ are contained within an envelop decreasing with $n$.

As a consequence, if the point $(\widehat{\lambda}_1,\widehat{\xi}_1)$ lies in the stability region, the points $\widehat{\lambda}_n<\widehat{\lambda}_1$ for $n\neq 1$ typically lie inside as well, as a result of the very shape of the shaded area.

The equation of the upper limit of the stability region reads \cite{PirizPRE2009},
\begin{equation}\label{eq:trian}
  \widehat{\xi} = 1-\sqrt{\widehat{\lambda}}.
\end{equation}
We thus need $\widehat{\xi}_1< 1-\sqrt{\widehat{\lambda}_1}$ for the system to be stable, which reads,
\begin{equation}\label{eq:stab0}
 \frac{\rho g \xi_1}{\sqrt{3}Y}  < 1-\sqrt{\frac{\rho g \lambda_1}{4\pi G}}.
\end{equation}
Following Ref. \cite{PirizPRE2009}, we introduce,
\begin{equation}\label{eq:p0}
    p_0=\rho g h,
\end{equation}
allowing to rewrite Eq. (\ref{eq:stab0}) as

\begin{equation}\label{eq:stab1}
 \frac{p_0}{\sqrt{3}Y}\frac{\xi_1}{h}  < 1-\sqrt{\frac{p_0}{4\pi G}\frac{\lambda_1}{h}}.
\end{equation}

We can consider here $\xi_1/h=\widehat{\rho}(1)/\widehat{\rho}(0)$. Using Eq. (\ref{eq:amplirelat}) for the parabolic time profile beam and $\lambda_1=2 \pi R$, we find

\begin{equation}\label{eq:stab2}
 \frac{p_0}{Y}\frac{\sqrt{3}}{\pi^2 N^2}  < 1-\sqrt{\frac{p_0}{2\pi G}\frac{ \pi R}{h}}.
\end{equation}

If $\pi p_0 R/2 G h \ll 1$, which seems to be the case in realistic situations, we just have,

\begin{equation}\label{eq:stab3}
Y>\frac{\sqrt{3}p_0}{\pi^2N^2}
\end{equation}
 which is the criterion obtained in \cite{PirizPRE2009}.

\section{Canceling the first harmonic}\label{sec:cancel}
An important issue for the present problem is the presence of the first harmonic. One reason is that its amplitude is usually the largest one. But most of all, the RTI analysis we used has been developed for a planar interface. If the interface is circular, the planar analysis is still valid providing the wavelength of the perturbation is much smaller than the circumference of the circle. In the present case, the first excited wavelength $\lambda_1$ is precisely the circumference. It is thus probable that the planar analysis fails for this mode. Canceling the first harmonic could solve the two problems at once: the planar analysis would just be applied to the next modes $\lambda_1/2$, $\lambda_1/4\ldots$, with a better accuracy and also, we could reduce significantly the irradiation asymmetry.

From Eq. (\ref{eq:fini1}), we see than canceling the first harmonic means tailoring the pulse shape $I(t)$ and the rotation frequency $\Omega$ to fulfill,
\begin{equation}\label{eq:1st harmo}
\int_{-T/2}^{T/2} I(t)\cos(\Omega t)dt=0,
\end{equation}
where we choose $t=0$ as the middle of the pulse. Substituting $u=\Omega t$ gives
\begin{equation}
\int_{-\Omega T/2}^{\Omega T/2} I(t/\Omega)\cos(u)du=0.
\end{equation}
Since $I$ is necessarily positive, the condition cannot be fulfilled if the cosine function takes only positive values in the integration range, i.e. $\Omega T/2<\pi/2$. We thus have a requirement on the beam duration $T$, with $T>\pi/\Omega$. Practically, this means the beam must rotate more than an half round,a condition which should be met obviously. With $\Omega$ in the GHz range, this imply a beam pulse longer than a few nanoseconds.

\begin{figure}[t]
\includegraphics[width=0.8\textwidth]{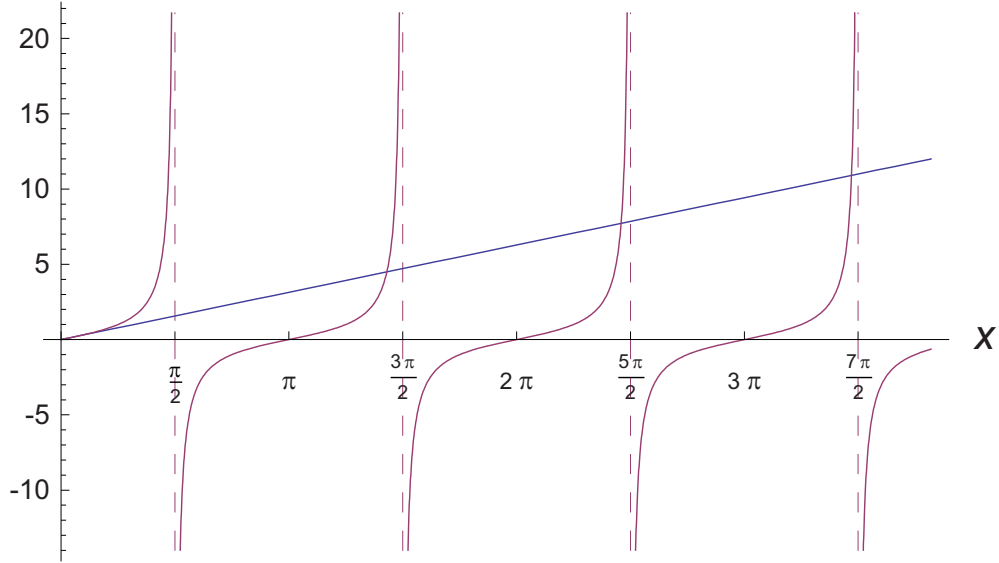}
\caption{Plots of functions $x$ and $\tan x$ vs. $x=\Omega T/2$. The equation $x=\tan x$ has an infinite number of solutions.} \label{fig:magic}
\end{figure}

Besides this requirement, the quadrature (\ref{eq:1st harmo}) is an oscillating integral vanishing for $\Omega\rightarrow\infty$. If it does so oscillating around 0, there will be an infinite number of values of $\Omega T$ canceling the first harmonic. Consider the parabolic profile introduced by Eq. (\ref{eq:para}). The first harmonic amplitude $\widehat{\rho}(1)$ is given setting $n=1$ in Eq. (\ref{eq:harmopara}). Imposing $\widehat{\rho}(1)=0$ gives,
\begin{equation}
\frac{\Omega T}{2}=\tan\frac{\Omega T}{2}.
\end{equation}
These two functions are plotted on Figure \ref{fig:magic}, and one can check the equation above has an infinite number of solutions. The first ones can be found numerically. Then, the solutions are found just before $x_k=(2k+1)\pi/2$, with $k\in \mathbb{N}$. Considering $\tan x\sim (x_k - x)^{-1}$ near $x_k$, we can find an approximation for $x_k$. The ``magic values'' of the product $\Omega T/2$ canceling the first harmonic are eventually given by,
\begin{eqnarray}
  x_1 &=& 4.493, \nonumber\\
  x_2 &=& 7.725, \nonumber\\
  x_3 &=& 10.904, \nonumber\\
      &\cdots&  \nonumber\\
  x_k &= &  (2k+1)\frac{\pi}{2}-\frac{1}{(2k+1)\pi/2}+\mathcal{O}(1/k^2).
\end{eqnarray}
By choosing any of these $\Omega T=2x_k$, the main asymmetry will come from the second harmonic only. The present procedure can be adapted to any beam profile and should allow to determine the optimum stability setup accordingly.

\section{Conclusion}
We have conducted the Fourier analysis of the ion density deposited on the circular absorber by a rotating beam. For a point-like beam with temporal profile, calculation can be performed exactly, and the spectrum of the deposited ion density expressed in terms of the spectrum of the temporal profile. The extension of the calculation to a spatially extended beam can also be assessed exactly, and a geometrical form factor is found correcting the point-like spectrum. As long as the beam radius is smaller than the absorber radius, the 1D spectrum, accounting only for the beam temporal profile, can be used for 2D beams. A stability analysis elaborating on the results from Piriz \emph{et al.} then gives the same result.

It has also been proved that it is perfectly possible to cancel the first harmonic of the beam deposition around the absorber. This allows to further reduce the irradiation asymmetry, and render more realistic the use of a RTI planar interface analysis. Calculations have been performed selecting a parabolic temporal beam profile, but they should eventually be conducted accounting for the beam profile chosen for the experiment.

\begin{acknowledgments}
This work has been supported by
Project ENE2009-09276, of the Spanish Ministerio de Educación y Ciencia,
Project  PAI08-0182-3162 of the Consejería de Educación y Ciencia de la Junta de Comunidades de Castilla-La Mancha,
and by the BMBF of Germany.
\end{acknowledgments}

\appendix

\section{Jacobian of Eq. (\ref{eq:real4})}\label{sec:ApA}
The Jacobian involved in Eq. (\ref{eq:real4}) has to do with the change of variables $(u,\theta_u)$ to $(R,\alpha)$. We start switching the $\theta$ origin with respect to Fig. \ref{fig6} so that $\theta=(\widehat{\mathbf{R}_c,\mathbf{R}})$. This does not bear any consequence on the present calculation because the differential elements remains the same by this simple transformation. Taking now the squares of the vectorial relations $\mathbf{R}_c+\mathbf{u}=\mathbf{R}$ and $\mathbf{u}=\mathbf{R}-\mathbf{R}_c$ gives,
\begin{eqnarray}\label{eq:Ap1}
  u &=& \sqrt{R^2+R_c^2-2RR_c\cos\theta}, \\
  \cos\theta_u &=& \frac{R^2-R_c^2-u^2}{2uR_c}. \nonumber
\end{eqnarray}
Inserting the value of $u$ from the first equation into the second one yields,
\begin{equation}\label{eq:Ap2}
    \cos\theta_u=\frac{R\cos\theta-R_c}{u}.
\end{equation}
The Jacobian matrix then reads,
\begin{equation}\label{eq:Ap3}
M_J=\left(
  \begin{array}{cc}
    \frac{\partial u}{\partial R} & \frac{\partial u}{\partial \theta} \\
    \frac{\partial \theta_u}{\partial R} & \frac{\partial \theta_u}{\partial \theta} \\
  \end{array}
\right).
\end{equation}
The determinant of this matrix is  found,
\begin{equation}\label{eq:Ap4}
J(\alpha, R)=\frac{R}{\sqrt{R^2+R_c^2-2RR_c\cos\theta}},
\end{equation}
where $u(R,\theta)$ as given by Eq. (\ref{eq:Ap1}) is at the denominator, which proves
\begin{equation}\label{eq:Ap5}
    u(R,\theta)J(R,\theta)=R.
\end{equation}
This result may not seems surprising. We basically switch from one system of polar coordinates to another with different origin. However, the fact that in such case the Jacobian is so simple is not so well known, and we choose to include the proof.

\bibliography{BibBret}

\end{document}